\documentclass[aip,reprint]{revtex4-1}
\usepackage[svgnames]{xcolor}
\usepackage[version=3]{mhchem}
\usepackage{graphicx}
\usepackage{blindtext}
\usepackage{siunitx}
\usepackage{miller}
\usepackage{xcolor}
\usepackage{float}
\usepackage{soul}
\begin{document}

\title{Epitaxial growth and thermodynamic stability of \ce{SrIrO3}/\ce{SrTiO3} heterostructures}

\author{D.~J.~Groenendijk}
\email{d.j.groenendijk@tudelft.nl}
\author{N.~Manca}
\author{G.~Mattoni}
\author{L.~Kootstra}
\affiliation{Kavli Institute of Nanoscience, Delft University of Technology, P.O. Box 5046, 2600 GA Delft, Netherlands.}
\author{S.~Gariglio}
\affiliation{Department of Quantum Matter Physics, University of Geneva, 24 Quai Ernest-Ansermet, 1211 Gen\`eve 4, Switzerland.}
\author{Y.~Huang}
\author{E.~van~Heumen}
\affiliation{Van der Waals-Zeeman Institute, Institute of Physics (IoP), University of Amsterdam, Science Park 904, 1098 XH, Amsterdam, Netherlands.}
\author{A.~D.~Caviglia}
\affiliation{Kavli Institute of Nanoscience, Delft University of Technology, P.O. Box 5046, 2600 GA Delft, Netherlands.}

\date{\today}


\begin{abstract}
Obtaining high-quality thin films of 5$d$ transition metal oxides is essential to explore the exotic semimetallic and topological phases predicted to arise from the combination of strong electron correlations and spin-orbit coupling. Here, we show that the transport properties of \ce{SrIrO3} thin films, grown by pulsed laser deposition, can be optimized by considering the effect of laser-induced modification of the \ce{SrIrO3} target surface. We further demonstrate that bare \ce{SrIrO3} thin films are subject to degradation in air and are highly sensitive to lithographic processing. A crystalline \ce{SrTiO3} cap layer deposited \textit{in-situ} is effective in preserving the film quality, allowing us to measure metallic transport behavior in films with thicknesses down to 4 unit cells. In addition, the \ce{SrTiO3} encapsulation enables the fabrication of devices such as Hall bars without altering the film properties, allowing precise (magneto)transport measurements on micro- and nanoscale devices.
\end{abstract}

\maketitle


The intriguing electronic structure of 5$d$ transition metal oxides arises from the delicate interplay between competing energy scales. Iridium compounds display a particularly large spin-orbit coupling (SOC) of the order of $0.4\;\mathrm{eV}$, which leads to the formation of novel $J_\mathrm{eff} = 1/2$ and $J_\mathrm{eff} = 3/2$ states~\cite{kim2009phase}. The combination of this strong SOC and slight lattice distortions has recently drawn attention to \ce{SrIrO3} as a promising candidate to realise topological (semi)metallic phases~\cite{xiao2011interface, carter2012semimetal, chen2015topological, chen2016topological, fang2016topological}. Perovskite \ce{SrIrO3} is a member of the \textit{Pbnm} space group, featuring two glide planes and a mirror plane which are crucial in determining its band structure~\cite{zeb2012interplay, liu2016strain}. At atmospheric pressure, \ce{SrIrO3} crystallises in a $6H$-hexagonal structure, while its perovskite form can be obtained by applying high pressure and temperature and subsequent quenching~\cite{longo1971structure}. This requires particular care due to the high volatility of iridium oxides and competition with other phases such as \ce{Sr2IrO4} and \ce{Sr3Ir2O7}~\cite{tian2013synthetic}. These extreme conditions can be avoided by resorting to thin film growth, where epitaxial constraint can be used to synthesize perovskite \ce{SrIrO3} films~\cite{kim2005metalorganic, jang2010pld, gruenewald2014compressive, matsuno2015engineering, zhang2015tunable, biswas2014metal, nie2015interplay, liu2015breakdown, liu2016strain, nishio2016thermodynamic}. \ce{SrIrO3} films are generally grown by pulsed laser deposition (PLD), where a relatively high oxygen pressure ($0.01$ - $1\;\mathrm{mbar}$) is required to control the \ce{Ir} oxidation state~\cite{nishio2016thermodynamic}. In such high pressure conditions, the interaction dynamics between the expanding plume and the background gas are very complex~\cite{amoruso2006propagation}. This can readily result in slight deviations from the ideal film stoichiometry, which can strongly affect the electrical properties through the formation of crystal defects. Electrical transport measurements of \ce{SrIrO3} films reported in literature show a rather large variability, which brings to question the role of disorder and secondary phase formation on the film properties~\cite{gruenewald2014compressive, matsuno2015engineering, zhang2014sensitively, zhang2015tunable, biswas2016emergence}.

In this Letter, we identify key issues related to the growth and stability of \ce{SrIrO3} thin films and study how these affect their electrical properties. First, we show that the morphology and stoichiometry of the \ce{SrIrO3} target surface is progressively modified by laser ablation, reducing the growth rate and affecting the electrical properties of thin films. We further demonstrate that the transport properties of \ce{SrIrO3} films are subject to degradation over time in ambient conditions, and that the films are highly sensitive to lithographic processing. These complications make it difficult to perform systematic and reproducible transport measurements. A crystalline \ce{SrTiO3} (STO) cap layer deposited \textit{in-situ} prevents film degradation and enables us to obtain metallic behavior in films with thicknesses down to 4 unit cells. In addition, the STO encapsulation preserves the film quality during lithographic processing, allowing the fabrication of Hall bars for precise (magneto)transport studies.

\begin{figure}[ht]
\includegraphics[width=\linewidth]{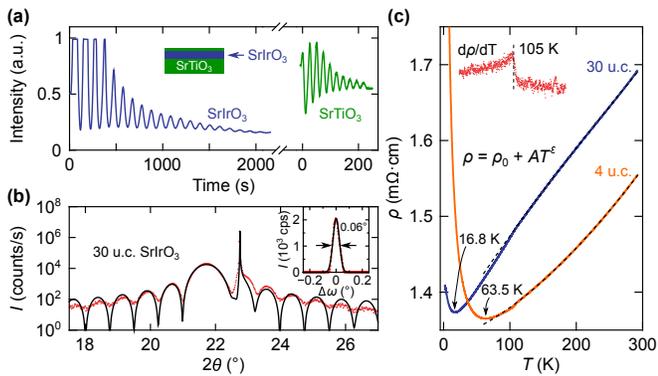}
\caption{\label{Fig1} (a) RHEED intensity oscillations of the specular spot during the growth of a 20 u.c.~\ce{SrIrO3} film with a 10 u.c.~STO cap layer. (b) X-ray diffraction scan around the \hkl(001) reflection (pseudocubic notation) of an STO-capped 30 u.c.~\ce{SrIrO3} film. Inset: rocking curve around the \ce{SrIrO3}\hkl(001) reflection. (c) Resistivity versus temperature for STO-capped \ce{SrIrO3} films with a thickness of 30 u.c.~(blue line) and 4 u.c.~(orange line). The dashed lines are fits to the data from room temperature to $105\;\mathrm{K}$. The inset shows $\mathrm{d}\rho/\mathrm{d}T$ of the 30 u.c.~film, where a change of the slope occurs at $105\;\mathrm{K}$.}
\end{figure}

\begin{figure}[ht]
\includegraphics[width=\linewidth]{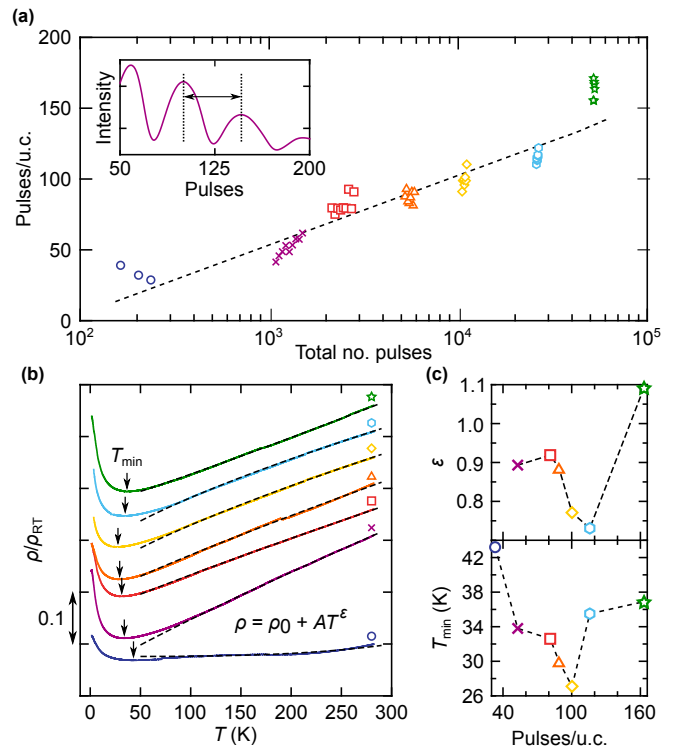}
\caption{\label{Fig2} (a) Pulses per u.c.~estimated from RHEED intensity oscillations during the growth of a series of 10 u.c.~\ce{SrIrO3} films. The inset shows the estimation of the number of pulses/u.c.~from the period of the  oscillations. (b) Resistivity versus temperature curves of the films shown in panel (a). The curves are rescaled to their room-temperature resistivity value and offset for clarity. (c) Parameters extracted from the $\rho(T)$ curves as a function of the average number of pulses/u.c.}
\end{figure}
\begin{figure}[ht]
\includegraphics[width=\linewidth]{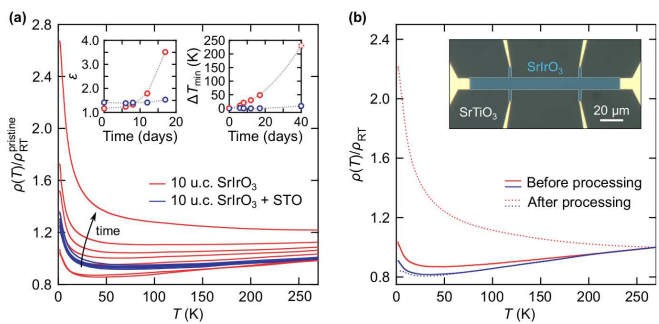}
\caption{\label{Fig3} (a) $\rho(T)$ for two 10 u.c.~\ce{SrIrO3} films, without (red line) and with (blue line) STO cap layer. The curves were measured for both samples simultaneously over a period of 40 days. The resistivity values are rescaled to the room-temperature resistivity of the measurement on day 0. The inset shows $\varepsilon$ and the variation of the upturn temperature $\Delta T_{\mathrm{min}}$ as a function of time. (b) $\rho(T)$ curves of 14 u.c.~films without and with an STO cap layer before (solid lines) and after (dashed lines) lithographic processing. Inset: optical image of a capped \ce{SrIrO3} film patterned into a Hall bar by \ce{Ar} dry etching.}
\end{figure}

\ce{SrIrO3} and STO films were deposited on commercially available \ce{TiO2}-terminated \hkl(001)\ce{SrTiO3} substrates (CrysTec GmbH) by PLD using a \ce{KrF} excimer laser (Coherent COMPexPro 205, \ce{KrF} $248\;\mathrm{nm}$)~\cite{kawasaki1994atomic, koster1998quasi}. An energy density of $1\;\mathrm{J}/\mathrm{cm}^2$ and a repetition rate of $1\;\mathrm{Hz}$ were used. The laser spot size was $2.5\;\mathrm{mm}^2$, corresponding to a dose of $25\;\mathrm{mJ}/\mathrm{pulse}$. The target was rotated during ablation such that the total ablated area forms a ring of $70\;\mathrm{mm}^2$. The incident angle of the laser on the target surface was $45^\circ$ and the target-substrate distance was set to $55\;\mathrm{mm}$. The depositions were performed in an oxygen pressure of $0.1\;\mathrm{mbar}$ and with a substrate temperature of $700^\circ\mathrm{C}$ as measured with an optical pyrometer. The growth conditions used in this work are comparable to those adopted most commonly in literature~\cite{gruenewald2014compressive, matsuno2015engineering, zhang2014sensitively, zhang2015tunable}. The relatively high oxygen pressure is required due to the noble metal character of \ce{Ir}, and causes the plume front to reach a stationary state after approximately $25\;\mathrm{mm}$\cite{nishio2016thermodynamic, amoruso2006propagation}. The diffusive propagation of the ablated species beyond this distance strongly reduces the deposition rate and can affect the film stoichiometry by preferential scattering of lighter elements with the background gas~\cite{wicklein2012pulsed}. In this regime, the kinetic energy of the species is of the order of the thermal energy, which is lower than the activation energy for surface diffusion of adatoms and can affect the growth mode~\cite{xu2013impact}. The growth was monitored by \textit{in-situ} reflection high-energy electron diffraction (RHEED). After growth, the samples were annealed in an oxygen pressure of $300\;\mathrm{mbar}$ for 1 hour and cooled down to room temperature in approximately 2 hours to compensate for possible oxygen deficiency. Single-crystal \ce{SrTiO3} and ceramic \ce{SrIrO3} targets were used. The \ce{SrIrO3} target was sintered in a sealed container at $950^\circ\mathrm{C}$ for 12 hours, followed by 24 hours at $1050^\circ\mathrm{C}$. Before use, the target surface was ground with fine grit sandpaper. X-ray diffraction data was acquired using a PANalytical X-PertPRO MRD equipped with a monochromator. Resistivity measurements on the samples shown in Figure~\ref{Fig1}(c) were performed in a Hall bar geometry fabricated by \ce{Ar} dry etching and e-beam evaporation of metal contacts, while the measurements shown in Figures~\ref{Fig2} and~\ref{Fig3} were performed in a van der Pauw configuration. The lithographic processing relied on the use of polymethyl methacrylate (PMMA) resists and standard chemicals such as acetone and isopropyl alcohol. 


In Figure~\ref{Fig1} we present the structural and electrical characterization of STO-capped \ce{SrIrO3} thin films deposited in optimum conditions. RHEED intensity oscillations, shown in Figure~\ref{Fig1}(a), monitor the growth rate and are observed for both the 20 u.c.~\ce{SrIrO3} film and the 10 u.c.~STO cap layer. Figure~\ref{Fig1}(b) shows an X-ray diffraction scan around the \hkl(001) reflection (pseudocubic notation) of an STO-capped 30 u.c.~\ce{SrIrO3} film. The clear finite size oscillations and the small FWHM of the rocking curve ($0.06^\circ$, Figure~\ref{Fig1}(b) inset) evidence long-range crystalline order. Additional X-ray diffraction measurements, such as reciprocal space mapping, show that the film is coherently strained on the substrate and are presented in the supplementary material~\cite{suppinfo}. Measurements over a larger $2\theta$ range indicate that, in these growth conditions, no secondary phases are formed~\cite{suppinfo}. The film thickness extracted from the fit ($12\;\mathrm{nm}$, solid line) is in good agreement with the number of unit cells estimated from RHEED and is confirmed by X-ray reflectivity~\cite{suppinfo}. The resulting $c$-axis parameter is $\SI{4.08}{\angstrom}$, which is consistent with the expected value taking into account the compressive strain from the STO substrate (+1.54\%) and imposing the conservation of the bulk unit cell (u.c.) volume.

Figure~\ref{Fig1}(c) shows the temperature dependence of the resistivity $\rho$ of two STO-capped \ce{SrIrO3} films with thicknesses of 30 u.c.~(blue line) and 4 u.c.~(orange line). In overall agreement with literature, the resistivity of the 30 u.c.~film decreases slightly with decreasing temperature, and displays a small upturn at low temperature~\cite{gruenewald2014compressive, matsuno2015engineering, zhang2014sensitively, zhang2015tunable, biswas2016emergence}. The small resistivity variation over the entire temperature range can be a signature of the semimetallic ground state reported in recent ARPES studies~\cite{nie2015interplay, liu2015breakdown}. Interestingly, we observe a slight change of slope at $T = 105\;\mathrm{K}$, which is reproducible across different samples and thermal cycles and is most likely related to the structural transition of the STO substrate from cubic to tetragonal phase~\cite{sato1985lattice}. This transition involves a rotation of the oxygen octahedra, shortening the in-plane lattice parameters and increasing the $c$-axis of the STO~\cite{loetzsch2010cubic}. Such cross-interface coupling has previously been observed for ultrathin correlated \ce{La_{1-x}Sr_xMnO3} (LSMO) films on STO, where a soft phonon mode, whose amplitude diverges at the STO phase transition, propagates into the atomic layers of the LSMO film and modifies its electronic properties~\cite{segal2011dynamic}. The detection of this slight structural distortion in the electrical transport of the \ce{SrIrO3} film is a fingerprint of the high quality of the interface, enabling the coupling between octahedral rotations of the substrate and the thin film. In addition, it demonstrates how sensitive the electric properties of \ce{SrIrO3} thin films are to octahedral rotations.

Despite its small thickness, the 4 u.c.~film still shows metallic behavior and has a resistivity comparable to the 30 u.c.~film. The resistivity shows an upturn at higher temperature, below which it increases up to approximately $2.4\;\mathrm{m\Omega\,cm}$ at $1.5\;\mathrm{K}$. Films with thicknesses below 4 u.c.~were found to display insulating behavior. The resistivity versus temperature data is fit from room temperature to $105\;\mathrm{K}$ by $\rho(T) = \rho_0 + AT^\varepsilon$ (dashed lines). Details regarding the data fitting are presented in the supplementary material~\cite{suppinfo}. In previous reports on \ce{SrIrO3} films, the temperature exponent $\varepsilon$ and upturn temperature $T_\mathrm{min}$ were considered as a measure of the film metallicity~\cite{biswas2014metal, biswas2016emergence}; here, we obtain $\varepsilon = 0.9$, $T_\mathrm{min} = 16.8\;\mathrm{K}$, and $\varepsilon = 1.6$, $T_\mathrm{min} = 63.5\;\mathrm{K}$ for the 30 and 4 u.c.~films, respectively. Literature reports of $T_\mathrm{min}$ for film thicknesses between $7$ and $35\;\mathrm{nm}$ vary from $25\;\mathrm{K}$ up to $175\;\mathrm{K}$~\cite{gruenewald2014compressive, matsuno2015engineering, zhang2014sensitively, zhang2015tunable, biswas2016emergence}. In the following, we will show that $\varepsilon$ and $T_\mathrm{min}$ are both affected by the progressive laser-induced modification of the target surface and film degradation in air over time.

To study how the growth and the electrical properties are affected by the modification of the target surface, a series of seven 10 u.c.~\ce{SrIrO3} films were deposited consecutively. The target was pre-ablated in deposition conditions with an increasing number of pulses between depositions to mimic the extended use of the target. Figure~\ref{Fig2}(a) shows the number of pulses per u.c.~extracted from the period of the RHEED oscillations as indicated in the inset. The entire set of oscillations is included in the supplementary material~\cite{suppinfo}. After 50,000 pulses incident on the rotating target, the number of pulses per u.c.~increases from 29 to 171. The reduction of the deposition rate can visually be recognized as an increased reflectance of the target surface and a progressive decrease of the plume size. Such a decrease in deposition rate has previously been reported for \ce{SrIrO3} and \ce{YBa2Cu3O_{7-x}} thin films and was related to stoichiometric and morphological changes of the target surface~\cite{jang2010pld, foltyn1991target}. We observe a similar change of target surface morphology by the formation of conical structures which align along the incoming laser direction, of which SEM images are included in the supplementary material~\cite{suppinfo}. This modification of the target surface was observed not only for $1\;\mathrm{J}/\mathrm{cm}^2$, but for different fluences ranging from 0.4 to $2.0\;\mathrm{J}/\mathrm{cm}^2$. These conical structures were previously shown to be \ce{Ir}-rich, indicating that the change in surface morphology is related to an \ce{Ir}-enrichment of the target surface~\cite{jang2010pld}. To corroborate this, we performed energy-dispersive X-ray (EDX) spectroscopy measurements on the target, finding a decrease of the \ce{Sr}/\ce{Ir} ratio of about 5\% after 240 pulses incident on the same site\cite{suppinfo}.

The resistivity versus temperature characteristics of the seven 10 u.c.~films are presented in Figure~\ref{Fig2}(b). The curves are normalized to their room temperature resistivity values and offset for clarity. Despite the significant decrease of the deposition rate, the resistivity and overall transport behavior of the thin films are comparable. The data is fit down to $105\;\mathrm{K}$ (dashed lines) to extract the temperature exponent $\varepsilon$. As shown in Figure~\ref{Fig2}(c), both $\varepsilon$ and $T_\mathrm{min}$ vary slightly and show a non-monotonic dependence on the deposition rate, displaying a minimum at approximately 100 pulses/u.c. We attribute this to an evolution of the film stoichiometry due to the interplay between target surface modification and preferential scattering of lighter species. As suggested by our EDX measurements, the laser ablation causes a progressive \ce{Ir}-enrichment of the initially stoichiometric target surface, resulting in a crossover from a \ce{Sr}- to \ce{Ir}-rich plasma plume as the number of pulses increases. The high background pressure can partially compensate for the incongruent ablation by preferential scattering of the lighter \ce{Sr} atoms, as has previously been observed for the PLD growth of homoepitaxial STO~\cite{wicklein2012pulsed}. In this picture, the minima in Fig~\ref{Fig2}(c) are indicative of a film with near-ideal stoichiometry, and the increase of $\varepsilon$ and $T_\mathrm{min}$ on either side of the minima demonstrates that a slight unbalance in the \ce{Sr/Ir} ratio can directly affect the electrical properties of \ce{SrIrO3} thin films. Films with optimum electrical properties can be obtained reproducibly by grinding the target surface prior to deposition and performing an in-situ pre-ablation in the growth conditions with a fixed number of pulses.

We further found that bare \ce{SrIrO3} films are subject to degradation in ambient conditions. Over time, films suffer from a loss of conductivity and shift towards insulating behavior. We monitored the progressive degradation of the transport properties by measuring resistivity versus temperature characteristics of a bare 10 u.c.~\ce{SrIrO3} film over the course of 40 days. In the time between measurements, the film was stored in ambient conditions. Figure~\ref{Fig3}(a) shows the $\rho(T)$ characteristics measured at different times (red lines) from which we extracted the time evolution of $\varepsilon$ and $T_\mathrm{min}$. The change from metallic to insulating behavior is reflected by an increase of the temperature exponent $\varepsilon$ and of $\Delta T_\mathrm{min}$, defined as $T_\mathrm{min}(t) - T_\mathrm{min}(t = 0)$ (Fig~\ref{Fig3}(a) inset). After 40 days, the metallicity is completely lost: the resistivity monotonically increases with decreasing temperature, and the curve can no longer be fit to the power law behaviour to extract the temperature exponent. We note how different $\rho(T)$ characteristics and values of $\varepsilon$ and $T_\mathrm{min}$ resemble the scatter of data reported in literature so far; it is thus possible that film degradation is one of the origins of their large variability.

A more abrupt change of the transport properties was observed when standard lithographic processing was attempted (as described in Methods). Figure~\ref{Fig3}(b) shows the relative variation of the resistivity versus temperature characteristic of a bare thin film (red lines), where the metallic behavior is completely lost after processing. The inability to pattern samples poses a serious roadblock to quantitative magnetotransport characterization of thin films, for which devices such as Hall bars are required.

We prevented the degradation of \ce{SrIrO3} films by the \textit{in-situ} deposition of a crystalline STO cap layer with a thickness of 15 unit cells. The addition of this cap layer preserves the electrical properties over time and enables lithographic processing. Figure~\ref{Fig3}(a) and (b) show the striking difference for the capped film (blue lines), where the $\rho(T)$ characteristics remain unchanged and $\varepsilon$ and $T_\mathrm{min}$ are approximately constant. It is possible that chemical decomposition occurs when bare films are stored in air or when lithography is performed, however no further studies have been performed on this issue.

We also found that encapsulation with amorphous STO yields stable films with no significant differences in their transport behavior, indicating that the choice of cap layer is not limited to crystalline STO~\cite{suppinfo}. This flexibility is particularly interesting considering that a crystalline STO cap layer imposes structural constraints to the film and restores a broken inversion symmetry, which could potentially affect its electrical properties. We did not observe significant differences in their $\rho(T)$ behavior, yet more detailed measurements are required to understand whether the electronic structure of the \ce{SrIrO3} films is affected by the presence of a crystalline STO cap layer.\\


In conclusion, we demonstrated the growth of high-quality epitaxial \ce{SrIrO3} thin films on STO substrates by PLD. Despite the continuous decrease of the growth rate due to the laser-induced modification of the target surface, the transport behavior of films deposited in different pre-ablation conditions was found to be comparable. The temperature exponent $\varepsilon$ and the upturn temperature of the films varied slightly, showing a non-monotonic dependence on the growth rate which we attributed to a crossover from \ce{Sr}- to \ce{Ir}-rich films. We further demonstrated how thin films suffer from degradation of their electrical properties in ambient conditions and after lithographic processing. The addition of an STO cap layer deposited \textit{in-situ} resulted in stable electrical properties over time and enabled us to measure metallic transport behavior in patterned films with thicknesses down to 4 unit cells. The sensitivity of the electrical properties of \ce{SrIrO3} thin films to slight deviations in stoichiometry and exposure to ambient conditions underlines the particular care that is required in their growth, characterization and processing.


\begin{acknowledgments}
This work was supported by The Netherlands Organisation for Scientific Research (NWO/OCW) as part of the Frontiers of Nanoscience program (NanoFront) and by the Dutch Foundation for Fundamental Research on Matter (FOM). 
\end{acknowledgments}

\bibliographystyle{apsrev4-1}
\bibliography{References}

\end{document}